\documentstyle[./psfig]{./mn11inch} 

\newcommand{\trise}{\mbox {$t_{\rm rise}$}}
\newcommand{\tplat}{\mbox {$t_{\rm plat}$}}
\newcommand{\tdecay}{\mbox {$t_{\rm decay}$}}
\newcommand{\phio}{\mbox{$\phi_{\rm obs}$}}
\newcommand{\thetao}{\mbox{$\theta_{\rm obs}$}}
\newcommand{\phij}{\mbox{$\phi_{\rm jet}$}}
\newcommand{\thetaj}{\mbox{$\theta_{\rm jet}$}}

\newcommand{\tpres}{\mbox{$\tau_{\rm pre}$}}
\newcommand{\tnu}{\mbox{$\tau_{\rm nu}$}}

\newcommand{\ngene}{\mbox{$N_{\rm gene}$}}
\newcommand{\pmutate}{\mbox{$P_{\rm mutate}$}}
\newcommand{\nchromosome}{\mbox{$N_{\rm chrom}$}}

\def\apgt{\ {\raise-.5ex\hbox{$\buildrel>\over\sim$}}\ }
\def\aplt{\ {\raise-.5ex\hbox{$\buildrel<\over\sim$}}\ }

\begin{document}

\title{
Precessing jets interacting with interstellar material as 
the origin for the light curves of gamma-ray bursts.}

\author[Portegies Zwart \& Totani]{
	Simon F. Portegies Zwart$^1$, 
	Tomonori Totani$^2$ \\ 
 $^1$ Massachusetts Institute of Technology, Cambridge, MA 02139, USA,
      Hubble Fellow \\ 
 $^2$ National Astronomical Observatory, Mitaka Tokyo 181-8588, Japan 
}
\date{Accepted 1993 December 11. Received 1993 March 17}
\maketitle




\begin{abstract}
We present an internal shock model with external characteristics for
explaining the complicated light curves of gamma-ray bursts.  Shocks
produce gamma-rays in the interaction between a precessing beam of
relativistic particles and the interstellar medium.  Each time the
particle beam passes the same line of sight with the observer the
inter stellar medium is pushed outward.  Subsequent interactions
between the medium and the beam are delayed by the extra distance to
be traveled for the particles before the shock can form.  This results
in a natural retardation and leads to an intrinsic asymmetry in the
produced light curves for gamma-ray bursts.  In addition we account
for the cooling of the electron-proton plasma in the shocked region,
which gives rise to an exponential decay in the gamma-ray flux. The
combination of these effects and the precessing jet of ultra
relativistic particles produces light curves which can be directly
compared with observed gamma-ray burst light curves.  We illustrate
the model by fitting a number of observed gamma-ray bursts which are
hard to explain with only a precessing jet.  With a genetic algorithm
we are able to fit several observed gamma-ray bursts with remarkable
accuracy. We find that for different bursts the observed fluence,
assuming isotropic emission, easily varies over four orders of
magnitude from the energy generated intrinsically. 

\begin{keywords}
	05 ---
	black hole physics --
	gamma-rays: bursts --
	gamma-rays: theory --
\end{keywords}

\end{abstract}


\section{Introduction}

Gamma-ray bursts (GRBs) are possibly among the most energetic event in the
Universe, emitting $10^{51-54}$ erg in soft gamma rays in an energy
range of $\sim$100\,keV--1\,MeV if the radiation is isotropic (see
e.g., Piran 1999 for a review).\nocite{Piran1999} They occur in
galaxies, at cosmological distances, and various theoreticians suggest
that the phenomenon is related to the death of massive stars, the
formation of compact objects or the coalescence between compact
objects.

The temporal structure of GRBs is characterized by a high variability
and durations vary from (possibly sub) millisecond to minutes (e.g.,
Norris et al. 1996)\nocite{1996ApJ...459..393N}, the origin of which
is not yet understood. Explanations range from multiple shock fronts
running into an ambient medium (Sari et al.\
1996),\nocite{1996ApJ...473..204S} expanding shells with brighter
patches and dimmer regions (Fenimore et al.\
1996)\nocite{1996ApJ...473..998F} to repeated series of pulses with
Gaussian or power-law profiles (Norris et al.\
1996).\nocite{1996ApJ...459..393N} However, a clear physical
explanation lacks and the geometry in these scenarios is not at all
clear; combinations of hydrodynamic, deteriation, expansion and
dilation time scales are introduced without a clear physical
explanation or understanding.  The light curves of GRBs can possibly
be explained by a beamed emission from a precessing accretion disk of
compact stars (see Portegies Zwart et al 1999; but also Roland et
al. 1994; Fargion 1998, 1999).\nocite{1994A&A...290..364R} Such a model
provides a rather simple physical explanation for the complex temporal
structure of GRBs.  Gamma-rays are emitted in a collimated cone or
beam, and precession of the jet produce multiple peaks in GRB light
curves when it passes the line of sight from an observer to the
central engine.  Portegies Zwart, Lee \& Lee (1999, hereafter
PZLL)\nocite{1999ApJ...520..666Z} investigated this scenario in detail
assuming that GRBs are produced by a particle emitting precessing jet
from the gamma-ray binary; a binary in which a neutron star fills its
Roche-lobe and transfers its mass to a black
hole (Portegies Zwart 1998).\nocite{1998ApJ...503L..53P} PZLL show
that the precessing jet scenario can roughly reproduce the observed
GRB light-curves by selecting reasonable precession parameters and a
distribution of the internal energy production within the jet.

There are, however, some difficulties with the physical picture of the
precessing jet as well as with fitting the model light-curves to the
observed GRBs.  A precessing jet tends to produce light curves in
which subsequent peaks appear at periodic intervals. The peaks
themselves may be slightly asymmetric but clearly lack the observed
strong asymmetries; individual peaks show a fast rise and a slow
decay.  Observed gamma-ray bursts also show no evidence for
periodicities at any time scale and Ramirez-Ruiz \& Fenimore (1999)
show that the width of the pulses in GRBs time histories remain
remarkably constant throughout the classic GRB phase.

The central engine cannot directly emit gamma-rays into a collimated
jet, owing to the well-known compactness problem (see e.g.\, Piran 1999).
The high opacity of $e^{\pm}$-pair creation reaction prevents
the free escape of the gamma-rays from such a compact region near the
central engine.  To overcome this difficulty, it is generally believed
that gamma-rays are emitted from shocked regions far from the central
engine, which are generated by ultra-relativistic outflows from the
GRB central engine with a Lorentz factor of typically $\Gamma \sim$ 
100--1000.  This problem could be solved if the jet carries
relativistic particles which shock the interstellar medium, far from
the central source.

We present a model for GRB light curves in which the above problems
are solved. The model reproduces observed GRB light-curves even better
than the older models of PZLL.  In this model gamma-rays are emitted
from outer shocked regions formed by interaction between the material
ejected ultra-relativistically in the precessing jet and ambient
interstellar medium. The compactness problem is then resolved
identically as in ordinary fireball models.  We also introduce a
retardation effect in the emission time and the moment the energy is
released in the form of gamma-rays.  This effect originates from the
finite velocity $<c$ of the relativistic particles and the slow
cooling of the shocked matter.

In section 2, we briefly review the GRB light-curves predicted by the
precessing jet scenario, and then present a detailed physical picture
and simple model of interaction between an ultra-relativistic jet and the
interstellar matter. In \S\,3 we describe the genetic fitting
procedure of the model to the observed GRB light-curves, and present
results for some GRB light-curves observed by the BATSE experiment.
We summarize in section\,4.

\section{Theory}

\subsection{Precessing jet}  
The model of the precessing jet was discussed in detail by PZLL (see
their section 3). For clarity we summarize the model and list its
parameters. For details concerning the energy generation process and
the origin of the precession we refer to PZLL.

The beam of relativistic particles emitted from the central object
precesses. As discussed by PZLL the direction in which the locus of
the precessing jet points is described by the precession angles
$\thetaj$ (the polar angle) and $\phij$ (the azimuth angle, both are
defined in a spherical polar coordinate system).  These angles are
obtained at any time $t$ from the initial direction in which the jet
points ($\theta^{\circ}_{\rm jet}$) and the precession and nutation
periods ($\tau_{\rm pre}$, $\tau_{\rm nu}$; see PZLL).

The direction from which an observer looks into the jet (two angles
$\thetao$ and $\phio$) now fixes the line of sight between the
observer and the central object.

The jet itself has a finite width. The intensity in the jet is assumed
to be a function of the distance to the central locus, as given by the
Blandford-Znajek (1977)\nocite{bz77} mechanism.  The intensity
distribution within the beam and the width of the jet can be described
with three dimension less parameters $\alpha$, $\beta$ and
$\delta$. Portegies Zwart, Lee \& Lee (1999) give a few examples of
the shape of the jet as a function of $\alpha$, $\beta$ and $\gamma$
in their figure 2. In these examples the width of the jet at half
maximum varies between $5^\circ$ (for $\alpha = 0.5$, $\beta=6$ and
$\gamma=0.4$) and $10^\circ$ ($\alpha = 0.6$, $\beta=6$ and
$\gamma=0.3$) with the maximum near $5^\circ$ in both cases.  More
extreme values for $\alpha$, $\beta$ and $\gamma$ have a stronger
effect on the location of the maximum and the width at half maximum
but the general shape remains very similar.  In the fitting performed
by PZLL, $\alpha$, $\beta$ and $\gamma$ were taken to be constant,
0.5, 6 and 0.3, respectively. Here they may vary.

With these parameters the width of the beam is of the order of
$\Gamma^{-1}$. And even if the initial width of the beam is smaller
than $\Gamma^{-1}$ sideways pressure guarantees an expansion of the
beam to approximately $\Gamma^{-1}$. The curvature of the jet front
should have an important effect on the observed shape of the GRB pulse
profile. The photons from a region off the line-of-sight are delayed
compared with those just on the line-of-sight, and this effect makes
the pulse profile as `fast rise and slow decay', as generally observed
for GRB pulses (Fenimore et al. 1996). For the simplicity we do not
include this effect in our calculation, but it should be noted that
this effect is similar to the effect of cooling of gamma-ray emitting
electrons, which is included in our model. Therefore, it can be argued
that the curvature effect is implicitly taken into account in our
model.

Here we must make an important assumption that the intrinsic time
variation of the relativistic energy output from the central engine is
much weaker than that caused by the precessing jet. The purpose of
this paper is to investigate whether the light curve of GRBs can be
understood mainly by the precessing jet effect, and it implicitly
assumes that the intrinsic time variability of the central engine is
rather smooth.  However, this assumption may not be true, and if the
central engine activity is also highly variable, our model does not
have strong predictability just like other stocastic explanations of
GRB light curves.  The intrinsic time variation of the relativistic
particle beam at its opening angle are chosen identical to the curve
adopted by PZLL, who construct time profiles for the central engine
from three components: an exponential rise with characteristic time
scale $\tau_{\rm rise}$, a plateau phase with time scale $\tau_{\rm
plat}$ and a stiff decay with time scale $\tau_{\rm decay}$ (see their
Eq. 11). This results in the following expression for the emitted flux
of relativistic particles in the direction of the observer as a
function of time:
\begin{equation}
	L_{\rm part}(t) = I(t, \trise, \tplat, \tdecay) 
	       J(\psi(\thetao, \thetaj, \phij), \alpha, \beta, \delta)
\label{Eq:Ljet}\end{equation}
where the angle $\psi$ between the observer $\hat r_{\rm obs}
(\thetao, \phio)$ and the central locus of the jet $\hat r_{\rm jet}
(\thetaj, \phij)$ is given by $\psi = \cos^{-1}(\hat r_{\rm
obs}\cdot\hat r_{\rm jet})$.

\subsection{Production of gamma-rays}
Two possible scenarios are known to generate the ultra-relativistic
shock required to explain GRBs: the external shock generated by the
collision between relativistic outflow with the ambient interstellar
matter, and internal shocks generated by collisions between
relativistic shells due to a relative velocity difference.  If the
relative velocity difference between internal shocks would be
stochastic, the complex time structure of GRB light curves is also
stochastic. This would make a detailed comparison between the light
curves of individual bursts with model calculations useless.  However,
we investigate the possibility that the light curves are determined by
precessing jets, which are deterministic; our model is not stochastic
and the energy generation by the central engine follows uniquely from
the provided parameters.

The precessing jet emits relativistic particles (it is not clear
whether the beam consists of an electron-positron or an
electron-proton plasma). The interaction between these and the
interstellar medium results in a deceleration of the particles and
their kinetic energy is transformed into gamma-rays.  This picture is
somewhat different than the external shock scenario, as in our case
the same region is shocked each time the precessing jet happens to
beam in the same direction.  The relativistic beaming makes gamma-rays
visible only from the shocked region on the line of sight to the
central engine.

The short precession and nutation times causes the precessing jet to
pass many times the line of sight, and hence jet energy is injected
with a complex and different time structure in each direction.  A
relativistic flow injected into one direction collides with other
flows which have been ejected earlier and slowed down by the
interaction with the interstellar medium.  This picture is therefore
somewhat intermediate between internal and external shock models.

We introduce two new effects to the light curves for the precessing
jet scenario discussed in PZLL: 1.) the time retardation caused by the
wake created by subsequent beam passages of relativistic particles and 2.) the
cooling of the shocked interstellar medium. The point 2 may also be
regarded as the pulse asymmetry generated by the curvature effect,
as discussed in the previous section.

The speed of the relativistic jet must be close to, but smaller than,
the speed of light to solve the compactness problem.  This results in
a time delay between the moment the particles leave the central engine
and the moment they collide with the interstellar medium.  While the
gamma-ray burst is active the location of the shocked region is pushed
farther away from the central engine by the relativistic particles.
The time delay will therefore build up in time.  This effect causes a
deviation from a periodic light curve produced by a precessing jet and
it extends the duration of the burst. The effect itself, however, does
not necessarily lead to a gradient in the pulse shape as a function of
time.

Gamma-rays are produced by some radiation process such as synchrotron
or inverse-Compton processes, as in the ordinary fireball
model. Colliding electrons in such processes have typical cooling
tails, which also affect the shape of the observable light curve.

The interaction of a precessing jet with the interstellar matter is a
complicated process and hard to model without detailed hydrodynamical
simulations with relativistic correction.  For our qualitative
analysis we rely on a simple prescription of the two discussed
effects, which may be sufficient.

\subsection{Implementation of the retardation time}

The kinetic-energy luminosity emitted by the precessing jet in
relativistic particles per steradian in the direction to the observer
is given by Eq.\,\ref{Eq:Ljet}.

For simplicity we assume that the Lorentz factor of the shock,
$\Gamma$, is constant. The total energy of the jet composed by
relativistic particles which is emitted in the
direction to the observer until time $t$ is then
\begin{equation}
E_{\rm part}(t) \equiv \int_0^t L_{\rm part}(t) dt \ .
\end{equation}

The deceleration radius $r_d$ is the distance from the central engine
at which the shock and thus the radiation are generated.  We assume
that $r_d$ is determined by the energy balance between $E(t)$ and the
energy carried by the mass of the interstellar matter which has been
swept up. This condition can be written as:
\begin{eqnarray}
4 \pi E_{\rm part}(t) &=& \frac{4 \pi}{3} r_d^3 n m_p \Gamma^2  \nonumber \\
r_d &=& 1.3 \times 10^{17} \left(\frac{E_{\rm part}
(t)}{\rm 10^{52} erg}\right)^{1/3}
n_1^{-1/3} \Gamma^{-2/3}_{100} \ \rm cm,
\label{Eq:rdec}\end{eqnarray}
where $n = n_1 \ \rm cm^{-3}$ is the interstellar matter density,
$\Gamma_{100} = \Gamma/100$ and $m_p$ is the proton mass.  As the
total energy $E(t)$ increases with time the deceleration radius
increases.

It takes time, $r_d/(1 - \Gamma^{-2})^{1/2}$, for the ejecta from the
jet to reach the shocked region at $r = r_d$.  (We use $c=1$.) The
emission from the shocked region is then delayed by a
\begin{eqnarray}
\Delta t &\sim& r_d \Gamma^{-2} / 2 \nonumber \\
&=& 2.1 \times 10^2  \left(\frac{E_{\rm part}
(t)}{\rm 10^{52} erg}\right)^{1/3}
n_1^{-1/3} \Gamma_{100}^{-8/3} \ \rm sec . 
\label{eq:delta-t}
\end{eqnarray}
If $t'$ is the time for the observer then $L_{\rm part, obs}(t')$ is the
observed light curve. The relation between $L_{\rm part, obs}(t')$ and the
light curve generated by the central engine $L_{\rm part}(t)$
is then given by
\begin{eqnarray}
t' = t + \Delta t(t)  \ , 
\label{eq:delta-t-solve} \\
L_{\rm part, obs}(t')dt' = L_{\rm part}(t) dt \ . 
\end{eqnarray}

Substitution of Eq. \ref{eq:delta-t} results in
\begin{eqnarray}
\frac{dt'}{dt} = 1 + 70 n_1^{-1/3} \Gamma_{100}^{-8/3}
E_{\rm part, 52}^{-2/3}(t) L_{\rm part, 52}(t) \ ,
\end{eqnarray}
where $E_{\rm part, 52} = E_{\rm part}/(10^{52}\rm erg)$ 
and $L_{\rm part, 52} = L_{\rm part}/(10^{52}\rm erg \
s^{-1})$.

\subsection{Implementation of the cooling time}

In order to predict the gamma-ray light curve from the particle
light curve defined above, we have to make several assumptions
regarding the gamma-ray production efficiency in the shocked region.
It is a standard view that electrons are accelerated efficiently
in the shocked region to radiate high energy gamma-rays. The
efficiency of electron acceleration, i.e., the fraction of
energy used to accelerate electrons in the total energy injected into
the shock, is not well known quantitatively, and here we assume
that it is constant during the GRB phenomenon. Then the light
curve of gamma-rays should be determined by taking into account
the cooling time scale of such accelerated electrons by gamma-ray
production.

We consider the cooling time of electrons assuming that the emission
process originates from electron-synchrotron radiation.  The energy
density of the shock heated matter in the rest frame of the shock is
$4n m_p \Gamma^2$ ($m_p$ and $m_e$ are the proton and electron
masses). Here we assume the magnetic field to be $B = (32 \pi \xi_B n
m_p \Gamma^2)^{1/2}$, where $\xi_B$ is the equipartition parameter for
the magnetic field and the energy density of the shocked matter
(Blandford \& McKee, 1976).

The photon energy $\varepsilon_\gamma$ of the electron synchrotron
radiation in the rest frame of the observer is given by
$\varepsilon_\gamma = \Gamma \gamma_e^2 e B / m_e$, where $\gamma_e$
is the Lorentz factor of the electrons measured in the rest frame of
the shock and $e$ is the electron charge.  Hence, we can estimate the
Lorentz factor of the electrons relevant to the synchrotron photons in
the BATSE range, as
\begin{eqnarray}
\frac{\gamma_e}{\Gamma} = 470 \Gamma_{100}^{-2}
\xi_B^{-1/4} n_1^{-1/4} \left( \frac{\varepsilon_\gamma}{\rm 100 keV}
\right)^{1/2} \ .
\end{eqnarray}
A typical value of $\gamma_e/\Gamma$ is 1 when the energy transfer
from protons into electrons is inefficient, while $\gamma_e/\Gamma
\sim m_p/m_e \sim$ 2,000 for efficient transfer.

What determines the typical photon energy range of GRBs is still a
difficult problem.  We do not discuss this and only use the above
estimate for the electron Lorentz factor to calculate the electron
cooling time.  In the rest frame of the shock the cooling time is
given by $t_{\rm rest} = 6 \pi m_e / (\sigma_T B^2 \gamma_e)$, and the
cooling time for the observer is then
\begin{eqnarray}
  t_{\rm cool, obs} = \frac{t_{\rm rest}}{2\Gamma} = 5.5 \times 10^{-2}
  \left( \frac{\varepsilon_\gamma}{\rm 100 keV} \right)^{-1/2}
  \Gamma_{100}^{-2} \xi_B^{-3/4} n_1^{-3/4} \ \rm sec \ .
\label{Eq:tcool}\end{eqnarray} 
This cooling time is convolved with the particle luminosity curve 
to get the observed photon light curve, i.e., 
\begin{eqnarray}
  L_{\rm \gamma, obs}(t) = \int_0^t dt' L_{\rm part, obs}(t') f_t(t, t')\ ,
\end{eqnarray}
where $f_t(t, t')$ is a transfer function:
\begin{eqnarray}
f_t(t, t') = \frac{\exp\left(- 
    \frac{t-t'}{t_{\rm cool, obs}}\right)}{t_{\rm cool, obs}} \ .
\end{eqnarray}

\section{Fitting of observed light curves}

The retardation effect and the cooling time increase the number of
free parameter to the model with the retardation time
(Eq.\,\ref{Eq:rdec}) and the characteristic cooling time scale $t_{\rm
cool}$ (Eq.\,\ref{Eq:tcool}). These parameters are functions of
$E(t)$, $\Gamma_{100}$ and $n_1$, which are not independent. For
clarity we assume $E = 10^{52}$ erg and $\Gamma_{100}$ is a few.

The model contains then a total of fourteen parameters
(Tab.\,\ref{Tab:param}) which are more or less free to be chosen
within the theoretical framework. For practical reasons we introduced
the dead time $\tau_{\rm dead}$ which is the initial time between the
BATSE trigger and the start-up of the real burst (see PZLL).  The
intrinsic parameters for the $\Psi$-dependence of the intrinsic
luminosity within the cone ($\alpha$, $\beta$ and $\delta$) are not
treated as fixed parameters, in contrast to PZLL, who assumed the
intrinsic luminosity distribution within the cone to be static between
bursts.


\begin{table}
\caption[]{
Model parameters which may vary per burst.
}
\bigskip
\begin{tabular}{lr} 
$\tau_{\rm dead}$ & initial epoch without signal \\
$\tau_{\rm rise}$ & start-up time of the central engine\\
$\tau_{\rm plat}$ & central engine plateau time \\
$\tau_{\rm decay}$& central engine decay time \\
$\tau_{\rm pre}$  & precession period \\
$\tau_{\rm nu}$   & nutation period \\
$\theta^{\circ}_{\rm jet}$& precession angle \\
$\theta_{\rm obs}$& observer angle \\
$\phi_{\rm obs}$  & observer angle \\ 
\\
$t_{\rm cool}$   & cooling time \\ 
$n_1$             & interstellar density \\
$\alpha$          & internal jet parameter \\
$\beta$           & internal jet parameter \\
$\delta$          & internal jet parameter \\ 
\end{tabular}
\label{Tab:param} 
\end{table}

\begin{table}
\caption{
Parameters for gamma-ray bursts BATSE 999, 1425, and 2067
}
\bigskip
\begin{tabular}{lrrr}
BATSE \#          &  999 & 1425 & 2067 \\ 
parameter & note \\ 
$\chi^2$          & 1.4         & 5.1       & 2.5         \\ 
$\tau_{\rm dead}$ & 1.00        & 0.883     & 1.328       \\
$\tau_{\rm rise}$ & 1.35        & 13.1      & 12.5      \\
$\tau_{\rm plat}$ & 7.50        & 7.36      & 0.860     \\
$\tau_{\rm decay}$& 0.705       & 0.25      & 3.63      \\
$\tau_{\rm pre}$  & 0.920       & 0.0452    & 0.214     \\
$\tau_{\rm nu}$   & 0.0047      & 0.193     & 0.216     \\
$\theta^{\circ}_{\rm jet}$&0.134& 0.0259    & 0.221     \\
$\theta_{\rm obs}$& 0.726       & 0.436     & 1.798     \\
$\phi_{\rm obs}$  & 3.23        & 0.0353    & 2.985     \\ 
\\
$t_{\rm cool}$    & 0.242       & 0.150     & 1.390     \\
$n_1$             & 1.60        & 0.050     & 0.107     \\
$\alpha$          & 0.25        & 0.207     & 0.530     \\
$\beta$           & 0.78        & 0.180     & 0.973     \\
$\delta$          & 0.0045      & 0.416     & 0.130     \\ 
\end{tabular}
\label{Tab:parameters}
\end{table}

The fits of PZLL are not all of the same high quality. Their
lightcurves still show symmetries, as could be expected from their
simple prescription of a precessing jet.  Possibly their fitting
algorithm was not robust enough to elude all the local minima in the
complex parameter space of the burst profile.  We therefore decided to
use a completely different fitting technique.

\subsection{The fitting procedure}\label{sect:fitting}

The first step in the fitting procedure is to determine the
background. This is done on the initial, $\sim 1800$, time bins of
64\,ms in the data stream of each observed burst. The average count
rate in this part is used as background.

We developed a genetic algorithm to elude the local minimum in the
optimization of the fitting process. This method performes considerably
better than the annealing technique adopted by PZLL, in finding the
global minimum and requires less iterations.

The array of parameters (see table\,\ref{Tab:param}) can be identified
with a chromosome in which each individual parameter (allele) is
identified as a gene, there are \ngene\, in each chromosome.  So, each
chromosome has a gene for the precession period \tpres\, the nutation
period \tnu\, etc. (see table\,\ref{Tab:param}).  Since the relative
effect of one parameter to the quality of the fit and the cross
correlation between the parameters is hard to determine we treat all
parameters independently.

The fitting process is started by selecting the allele for each gene
for the first chromosomes, which may contain a reasonable first guess
to the bursts profile.  This chromosome is copied \nchromosome\,
times, with $\nchromosome = \ngene+1$ if \ngene\, is uneven and
$\nchromosome = \ngene+2$ times otherwise.  The initially selected
chromosome is stored as potentially best fit.  All other chromosomes
are mutated to produce the genotype for the new populations.

The iterative process starts with a loop over the
following steps, which we call a generation:

For mutating a chromosome each gene has a probability \pmutate\, of
changing its value ($\pmutate\ \sim 5\%$).  The selected gene is then
incremented with 
\begin{equation}
0.5 \sqrt{12} \langle g_i \rangle (X_1+X_2-X_3-X_4),
\label{Eq:mutate}\end{equation}
where $\langle g_i \rangle$ is the mean value for gene $i$ of the
allele in all genotype, and $X_1$ to $X_4$ are four random numbers
between zero and unity.

The genotype are evaluated by calculating the $\chi^2$ of the models
with the observed gamma-ray burst.  For this purpose we calculate the
light curves associated with the selected chromosome and compare that
with the observed burst. The $\chi^2$ is then calculated.  (Note that
a 10 seconds gamma-ray bursts has more than 150 degrees of freedom.)

The fitness for each genotype $f_i$ is then calculated by rescaling the
$\chi^2$ for each burst via
\begin{equation}
	f_i = c_f + m_f \nu_i,
\end{equation}
where
\begin{eqnarray}
	\nu_i = \exp \left( a \chi_i^2/\langle \chi^2 \rangle \right), \\
	m_f = (b-1) {\langle \nu \rangle \over
                    \nu_{\rm max} - \langle \nu \rangle}, \\
	c_f = (\nu_{\rm max} - b \langle \nu \rangle) 
				 {m_f \over b-1}.
\end{eqnarray}
Here $a \sim -20$ and $b =2$ or 3 are constants, $\langle \nu \rangle$
and $\nu_{\rm max}$ are the mean and maximum value of $\nu_i$ in the
parental pool.  From the available genotype we randomly select
$\nchromosome$ parents.  The probability for parent selection is
proportional to its fitness: $f_i/\langle f \rangle$, where $\langle f
\rangle$ is the mean fitness in the parental pool. The genotype which
are fitter than average may be selected more than once and the less
fitter parents my stay unselected.

Each time two parents are selected they produce two offspring, which
populate the new phenotype.  Parents, however, are not allowed to be
identical: this would lead to two identical children.  Each gene from
one of the parents has a fifty-fifty chance to become part of one of
the two children.  In this way it is possible that two parents result
in two children with identical chromosomes as the parents, in which
case both children mutate (via Eq.\,\ref{Eq:mutate}).

If the chromosome with the lowest $\chi^2$ in the population appears
fitter than the potentially best fit, they are exchanged.  After
$N_{\rm gen}$ generations the chromosome with the lowest $\chi^2$ is
selected as the best fit.

\subsection{The fitting results}
\begin{figure}
(a)\psfig{figure=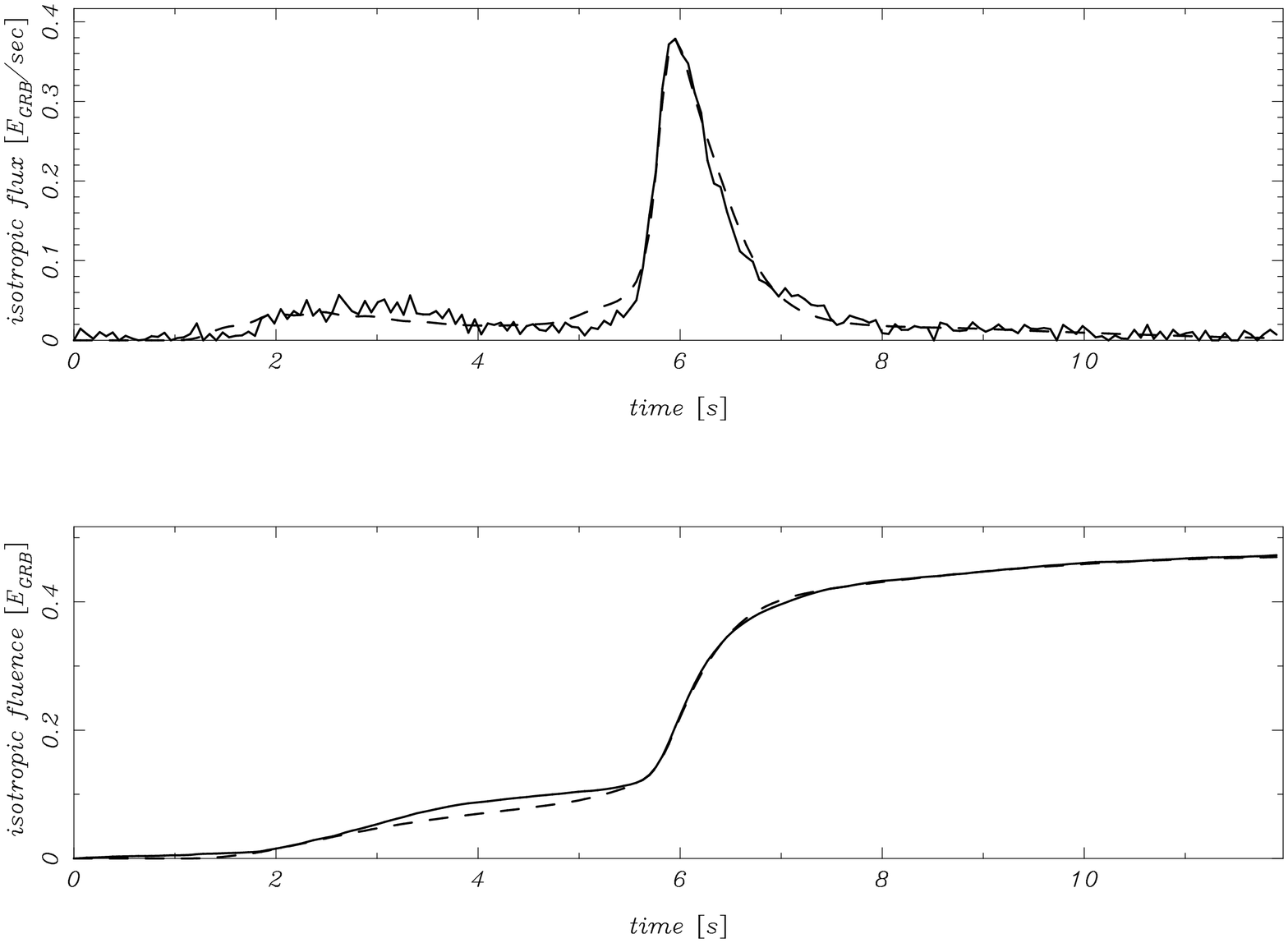,width=7.5cm,angle=-90}
(b)\psfig{figure=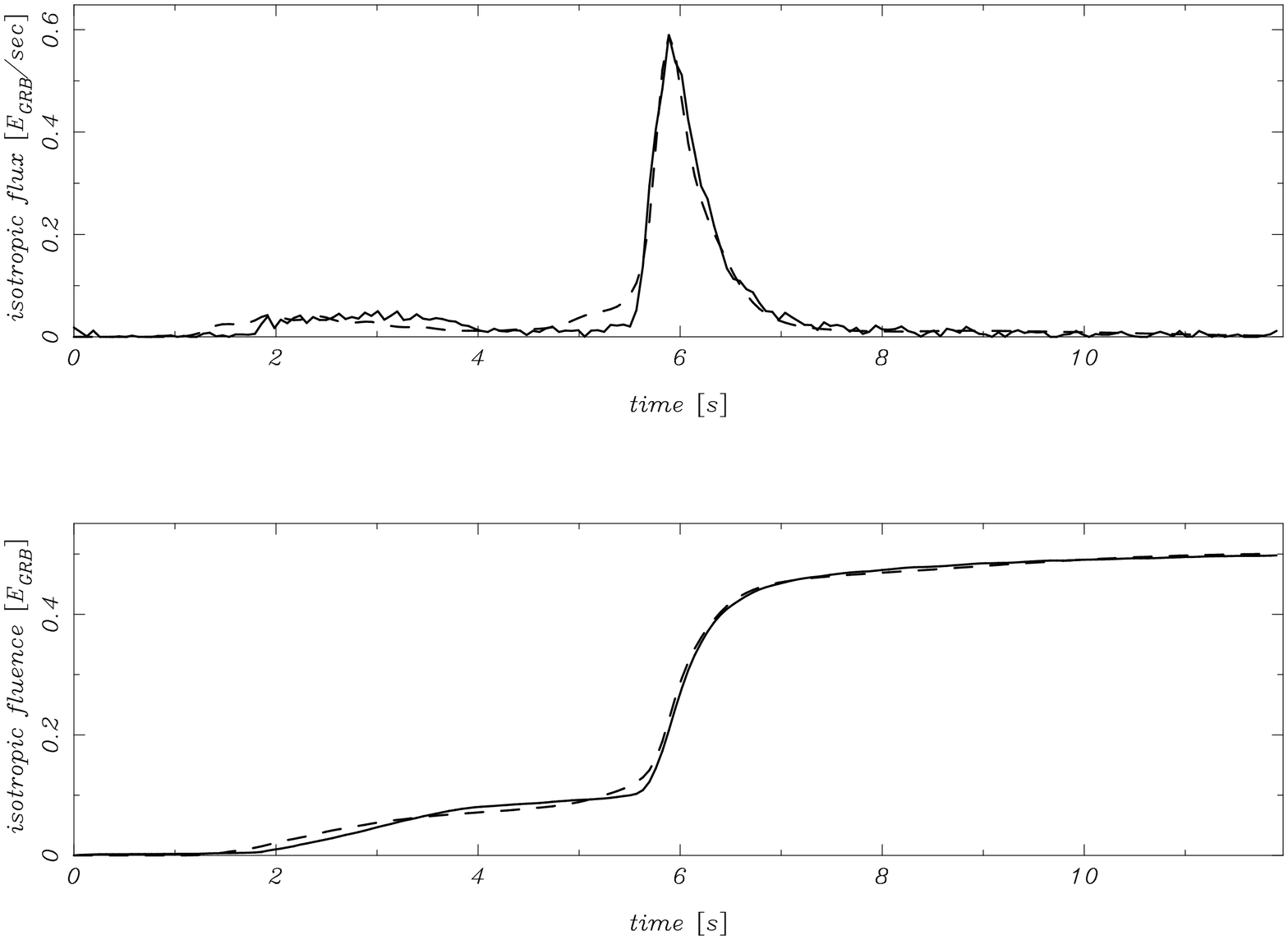,width=7.5cm,angle=-90}
(c)\psfig{figure=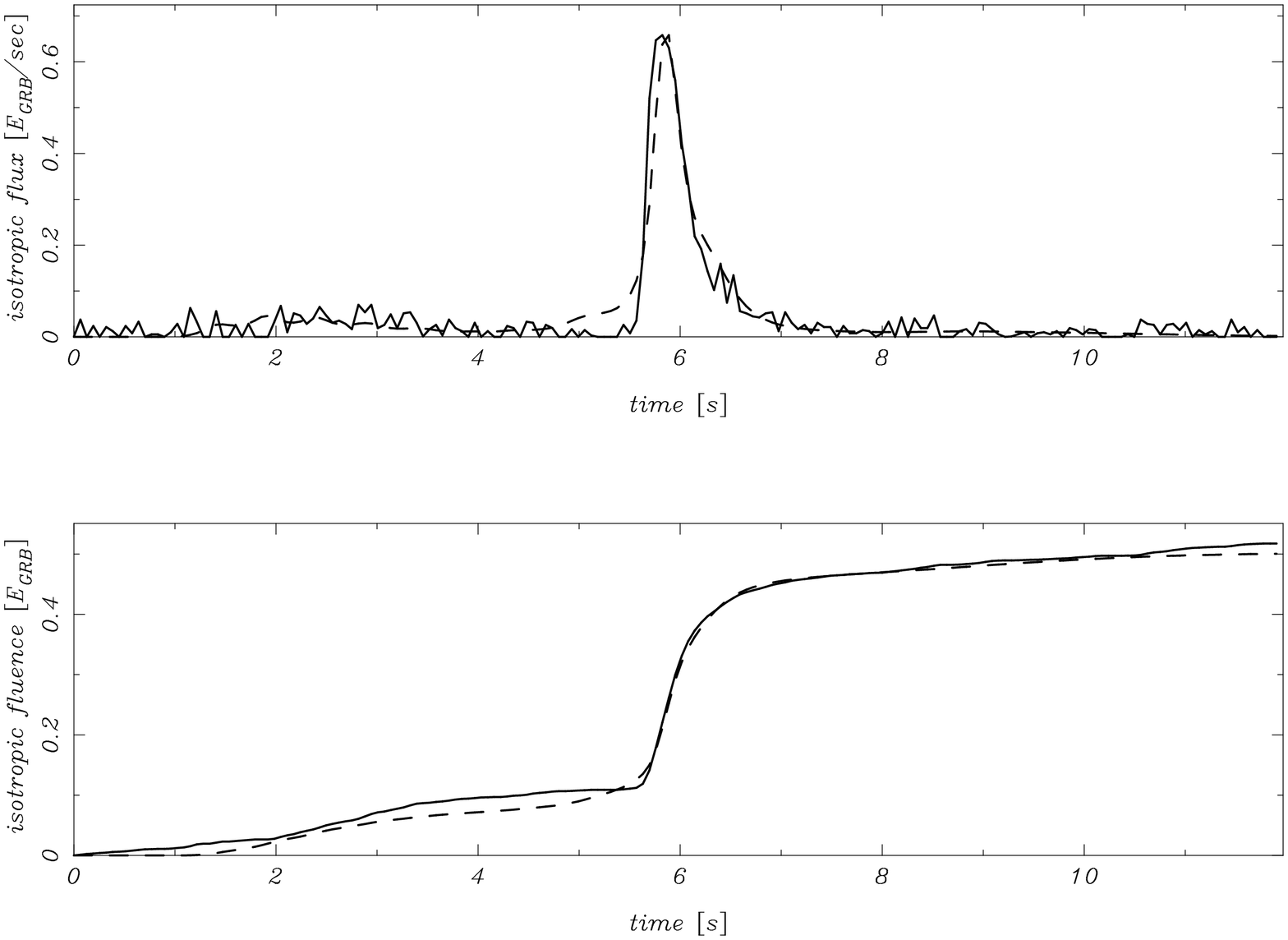,width=7.5cm,angle=-90}
\caption[]{ (a)Result of the genetic fitting to BATSE trigger number
999 (second energy channel: 50 -- 100 keV). Upper panel shows the real
burst profile (solid line) and the fit (dashed line) of isotropic
luminosity assuming isotropic radiation, where the lower panel shows
the same but for isotropic fluence. Here the isotropic luminosity and
fluence are given in units of the true luminosity and fluence of the
jet from the central engine integrated over all directions. Small
numbers of isotropic fluence indicate that we observe only a small
fraction of the true energy emitted as the jet.  Fit parameters are as
in Tab.\,\ref{Tab:parameters}.  (b) Fit to third energy channel of GRB
999 (100 -- 300 keV).  The obtained $\chi^2 = 1.29$, variable
parameters is $t_{\rm cool} = 0.1$. Other fit parameters are as in
Tab.\,\ref{Tab:parameters}.  (c) Fit to fourth energy channel of GRB
999 ($>300$\,keV).  The obtained $\chi^2 = 1.29$, $\delta$ as in
Fig.\,\ref{GRB999_II}, $t_{\rm cool} = 0.05$ (see
Tab.\,\ref{Tab:parameters} for other parameters).  }
\label{GRB999_I}
\label{GRB999_II}
\label{GRB999_III}
\end{figure}

\begin{figure}
(a)\psfig{figure=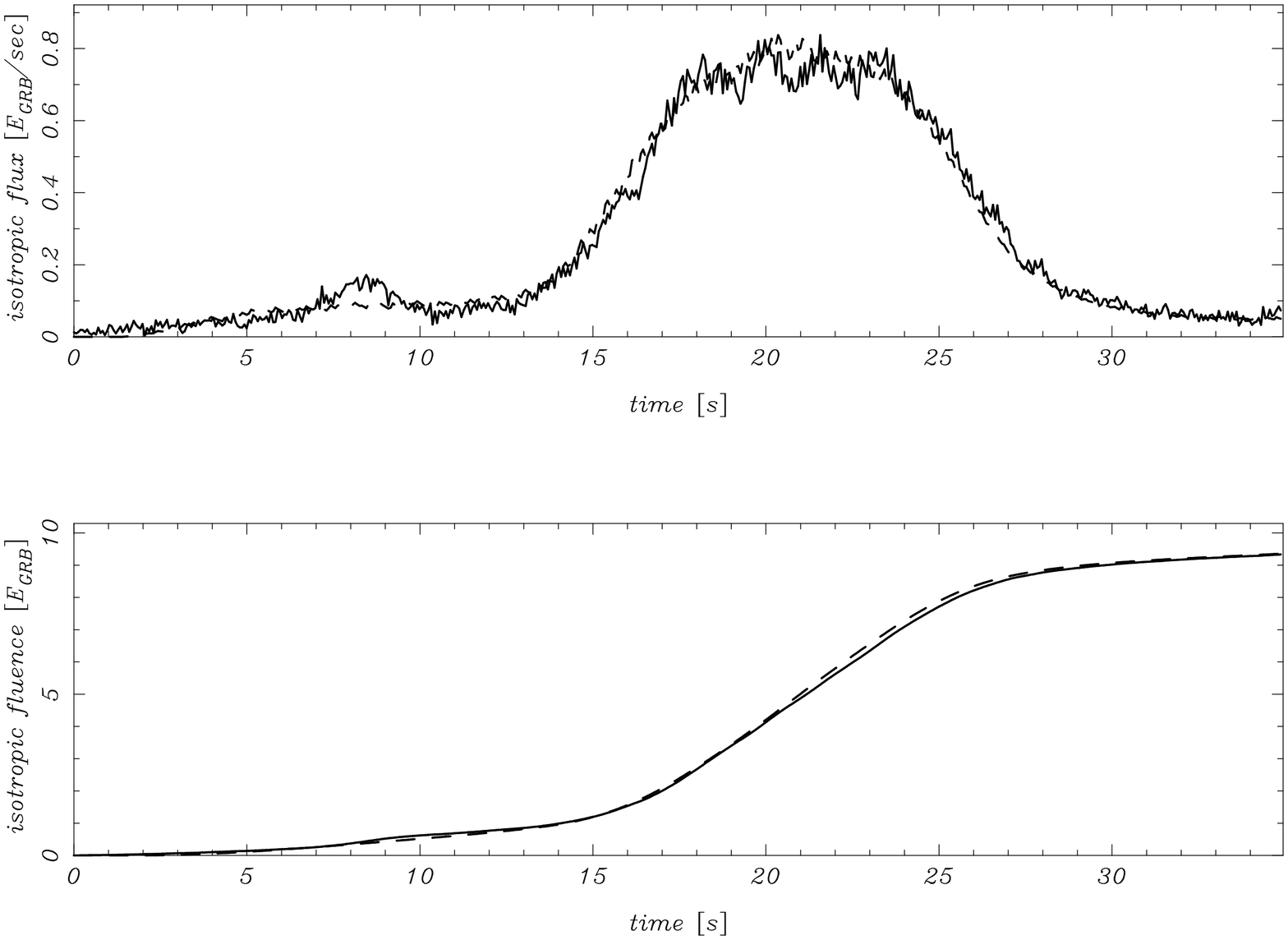,width=7.5cm,angle=-90}
(b)\psfig{figure=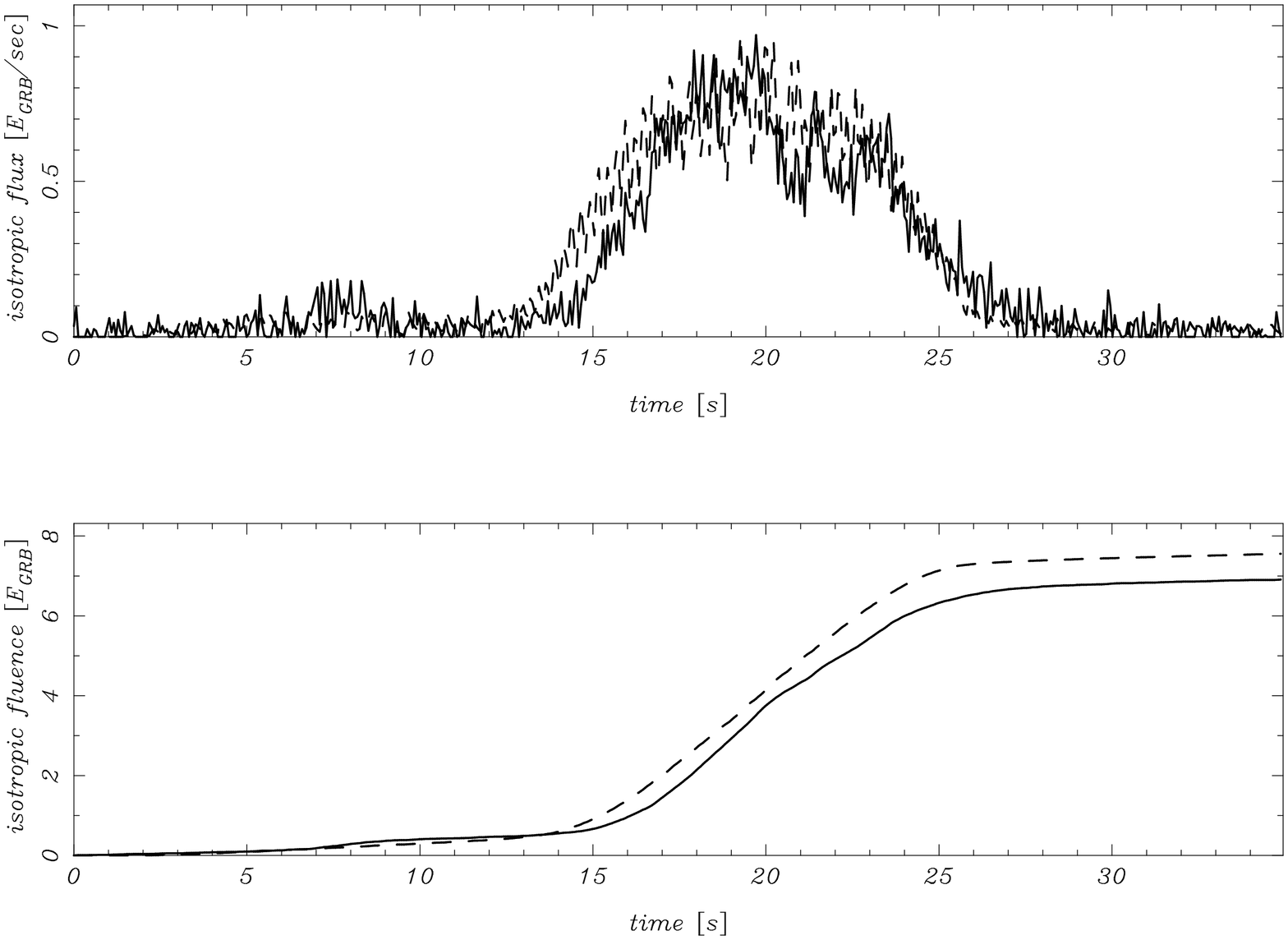,width=7.5cm,angle=-90}
\caption[]{ (a)Example to fit to second energy channel (50 -- 100 keV)
of GRB 2067.  The obtained $\chi^2 = 2.5$, $t_{\rm cool} = 1.39$ (see
Tab.\,\ref{Tab:parameters} for other parameters).  (b) Fit to third
energy channel of GRB 2067.  The obtained $\chi^2 = 14.4$, $t_{\rm
cool} = 0.15$ (see Tab.\,\ref{Tab:parameters} for other parameters).
}
\label{GRB2067_I}
\label{GRB2067_III}
\end{figure}

\begin{figure}
\psfig{figure=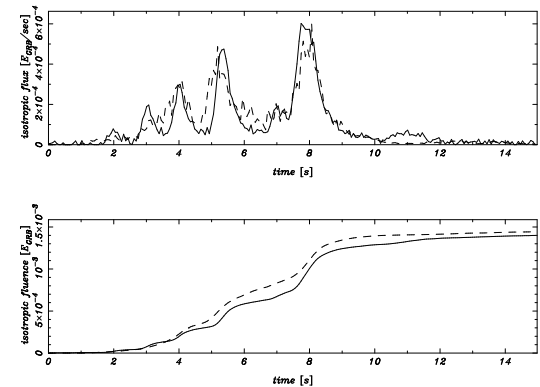,width=7.5cm,angle=-90}
\caption[]{
Fit to second energy channel of GRB 1425
(parameters are in Tab.\,\ref{Tab:parameters}).
}
\label{GRB1425}
\end{figure}

The model and fitting algorithm are tested on a number of observed
gamma-ray bursts. The result of the fits are presented in the figures
\ref{GRB999_I} to \ref{GRB1425}.

Most parameters in the model should be the same in different 
energy bands of the same burst.  The intrinsic precession and
nutation parameters, observers angles, the density of the interstellar
medium and the strength of the magnetic field should all be identical
for the same burst at different energy channels. Only the cooling time
of the interstellar medium and possibly the structure parameters of
the jet may differ between energy bands within the same burst.

Figures\,\ref{GRB999_I}a to \ref{GRB999_I}c illustrate how the same
jet parameters fit the different energy channels of an observed
gamma-ray burst. 
Only the cooling time for the interstellar medium
and one of the structure parameters of the jet are varied. 
The tail of the main pulse of this burst
becomes shorter in higher photon energy bands, and this trend is
well explained by the effect of the cooling. It should also be noted
that the similar trend is also expected by the curvature effect,
since the photons from regions off the line-of-sight are softened
by an effectively smaller Lorentz factor.

The $Y$-axis in these figures represents the fluence measured by an
observer assuming that the burst radiated isotropically, in units of
the true energy output from the central engine. Note that this number
may be larger as well as smaller than unity and shows a wide spread in
our fits.

Another example is presented in the figures\,\ref{GRB2067_I}a and
\ref{GRB2067_I}b, where we attempted to fit GRB trigger number 2067 in
two energy bands. The other energy bands are satisfactory fitted by
changing the cooling time and the internal structure of the jet.

Figure\,\ref{GRB1425} illustrates how the fit to GRB 1425 produces an
asymmetric burst due to the discussed effects, where PZLL had
difficulties explaining the increase in the time intervals between
peaks in GRB 1425.

\section{Discussion and conclusion}
The extension of the simple model (by Portegies Zwart et al. 1999) in
which a precessing jet is the cause of the complicated light curves
of gamma-ray bursts fits observed light curves with higher
quality. The increase in the model complexity causes an increase in
the required computer time per burst and makes the fitting much
harder.

We have obtained remarkable fits to the light curves of some real
gamma-ray bursts, by using a physical model of the precessing jet and
its interaction with the interstellar medium. Of course, there are
quite a few gamma-ray bursts for which we tried our fitting algorithm
but were unable to obtain satisfactory fits. However, it may be due to
the difficulty of the fitting and does not mean that the precession
jet scenario cannot explain all of the GRB light curves. Currently the
precession jet scenario is almost the only 'physical' or
'deterministic' explanation for the GRB light curves, and it deserves
further investigation. For clarity we assumed a simple and smooth 
light curve for the intrinsic luminosity of outflow from the central
engine, characterized by three time scales (initial rize, plateau,
and decay). The variation in the observed
output is in these cases caused by variations in the other model
parameters and the angle from which the source is observed.  Note,
however, that this choise is not a requirement and that the energy
output of each burst may well vary intrinsically.

We mainly consider the cooling of electrons as the origin of the
asymmetric shape of peaks in GRB light curves, but some other
explanations for the asymmetry are also possible.  One of these is the
delayed emission from shocked region off the line-of-sight to the
central engine (Fenimore et al. 1996).  In the external shock model of
GRBs, such an asymmetry is expected because gamma-rays emitted from
off-axis region will be delayed and softened compared with the
emission from the on-axis region, because of a geometric effect and
lower effective Lorentz factor of relativistic motion. A similar
effect is also expected in our model; when a precessing jet passes the
line-of-sight from an observer to the central engine, the emission
from on-axis region is promptly observed by us, but emission before
and after the passing beam should be delayed and softened because it
is from off-axis shocked regions generated by the precessing
jet. The intrinsic curvature of the jet front may also be important,
since the width of the jet in our model is comparable with the
visible size of the emission region ($\theta \sim \Gamma^{-1}$).
Although this effect is difficult to include exactly in our
calculation, we note that it is taken into account, at least
qualitatively.  We have fitted the GRB light-curves with the cooling
time as a free parameter, and this is not necessarily to be the
cooling time for electrons.  Off-axis emission is expected to show a
qualitatively similar effect, and it is possible that the cooling time
which we obtained in the fits may be the effect of off-axis emission.

It is known that the observed total energy emitted from GRBs assuming
isotropic radiation shows a wide dispersion by about 2-3 orders of
magnitudes.  Our fitting results also show a wide dispersion in the
isotropic energy by more than 3 orders of magnitudes, with a constant
true energy emission as a jet from the central engine. (See isotropic
fluences in Figs. 1, 2, and 3.) The model therefore provides a
possible explanation for the observed wide dispersion in the isotropic
total GRB energies.

\bigskip\noindent{\bf Acknowledgments} We thank Douglas Heggie for
speeding up the code.  This work was supported in part by the Research
for the Future Program of Japan Society for the Promotion of Science
(JSPS-RFTP97P01102), and by NASA through Hubble Fellowship grant
HF-01112.01-98A awarded to SPZ by the Space Telescope Science
Institute, which is operated by the Association of Universities for
Research in Astronomy, Inc., for NASA under contract NAS\, 5-26555.
Calculations are performed on the SGI/Cray Origin2000 supercomputer at
Boston University.



\end{document}